\documentstyle[11pt,moriond,psfig]{article}

\def\Journal#1#2#3#4{{#1} {\bf #2}, #3 (#4)}

\def\APJ{{\em Ap. J.}}
\def\APJS{{\em Ap. J. Suppl.}}
\def\AA{{\em A.\&A.}}
\def\MNRAS{{\em MNRAS}}

\def\be{\begin{equation}}
\def\ee{\end{equation}}
\def\bea{\begin{eqnarray}}
\def\eea{\end{eqnarray}}

\begin{document}
%\vspace*{4cm}
\title{AGN Models:  High-Energy Emission}

\author{Karl Mannheim}

\address{Universit\"ats-Sternwarte, Geismarlandstra{\ss}e 11, D-37083, Germany}

\maketitle

\abstracts{Astrophysical models for the high-energy emission of
blazars are reviewed.  Blazars ejecting relativistic radio jets at small
angles to the line-of-sight
are the only type of 
active galactic nuclei (AGN) discovered above 100~MeV.
The $\gamma$-rays apparently originate in the jets which explains
the    absence of $\gamma$-ray
pair creation attenuation. 
The  bulk Lorentz factor of the radio jets  is much smaller than
the Lorentz factors of the emitting particles requiring 
{\em in situ} particle acceleration such as Fermi acceleration
at shocks.   The mounting evidence for a
correlation between the optical and $\gamma$-ray emission argues for
the same  accelerated
electrons emitting the polarized optical
synchrotron photons to be  responsible
for the high-energy emission.
Alternatively, the $\gamma$-rays could be due to
electrons of a secondary origin related 
to the energy losses of  protons accelerated at
the {\em same} shocks as the   synchrotron 
emitting electrons.
In this case blazars would produce an observable flux of high-energy
neutrinos.
Unified schemes for AGN  predict a 
circum-nuclear warm dust torus 
attenuating
$\gamma$-rays above $\sim 300$~GeV emitted from within  
a central sphere of radius
$\sim 2\times 10^4r_{\rm S}$ which rules out
external Compton scattering models as the origin of
the TeV $\gamma$-rays from Mrk421 and Mrk501.
}

\section{Introduction}
The EGRET discovery of $\gamma$-ray sources at high galactic
latitudes and their identification with 
blazars ($\sim$ flat-spectrum radio-loud AGN) has prompted a large number of authors to think about possible explanations
for the phenomenon.  As of 1996, I have counted almost one model or one model
ramification per source scanning through NASA's Astrophysics Data System.  The
considerable theoretical interest documents the importance of the discovery and
at the same time demonstrates the on-going struggle for a viable explanation.
Of course, not all of the papers can be discussed in this review, and I have
decided to critically discuss only the basic model assumptions 
and their consequences.  The assumptions that (i)
blazars represent accreting supermassive
black holes  ejecting bipolar jets along
the angular momentum axis of the inner accretion disk and that (ii)
the $\gamma$-rays originate in the  
jets 
are common to all the models.  This 
consensus is based mainly on the facts that
EGRET has observed $\gamma$-rays only from AGN with jets
and that jets provide a natural resolution of the so-called 
{\em compactness problem}:
For $\gamma$-rays of energy $m_{\rm e}c^2$
the pair creation optical depth   $\gamma+\gamma\rightarrow e^++e^-$
of an Eddington accreting black hole is the universal constant
\begin{equation}
\tau_{\gamma\gamma} \simeq 
  {L_{\rm edd}\over 4\pi r_{\rm S} m_{\rm e}c^3}{3\sigma_{\rm T}
\over 16} =
{3\over 32} {m_{\rm p}\over m_{\rm e}}\simeq 200 
\end{equation}
\begin{figure}[t]
\vskip-2.5cm
\centerline{\psfig{figure=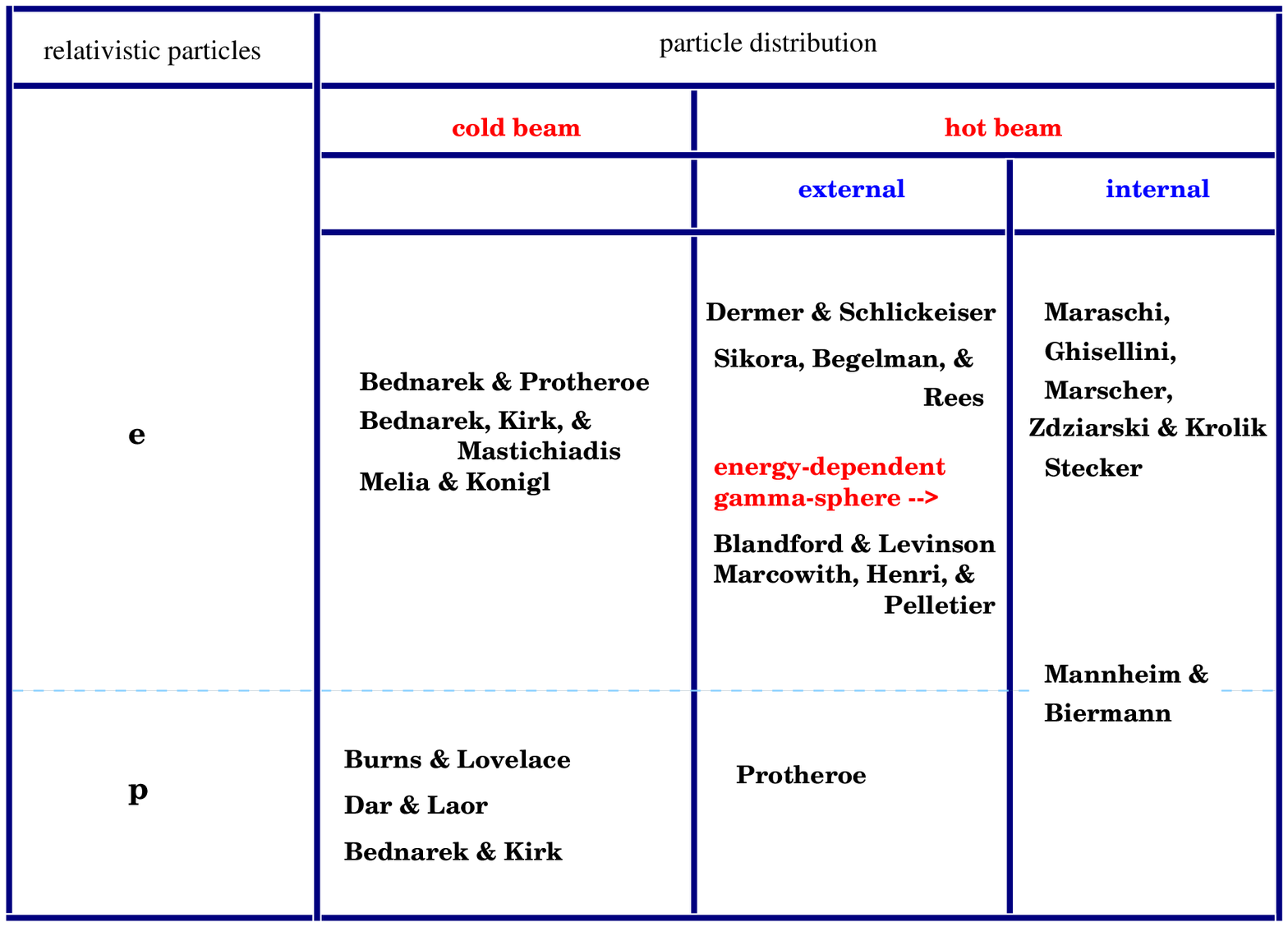,width=17cm,height=11cm,angle=-90}}
\caption{Coarse overview of models.  See Sect.~1.1 for details.}
\end{figure}
Eddington accretion onto a supermassive
black hole is the only mechanism efficient enough to explain
the high luminosities and short variability time scales of blazars
assuming isotropic emission.
However,
the large optical depth Eq.(1) is  inconsistent with the observed
optically thin power law spectra\cite{k97} and therefore
the emission must be anisotropic.

The accreting black hole picture for blazars is based
on  
unified schemes for AGN which are discussed extensively
in the contribution by Padovani\cite{p97}.
As a general cautious remark I want to emphasize that
the predictions of the
unified schemes for the density of matter and radiation in
the medium surrounding the jets are subject to change in
the course of further astronomical studies and should therefore
be regarded as preliminary.  Furthermore,
the unified schemes leave it essentially open whether
the jets form as a hydromagnetic $e^-p$ wind from the relativistic inner
accretion disk\cite{c86} or as a Poynting flux driven $e^\pm$ wind
extracting rotational energy from the central black hole\cite{bz77}.
The merging history of ellipticals as the typical
host galaxies of radio-loud AGN  lends some support for the
rotating black hole picture\cite{wc95}, although jets are also
associated with young stars which do not contain black holes.
It is hoped that the $\gamma$-ray properties of blazars probing
the physical conditions in the jets ameliorate theory in this important
issue.  However, baryon pollution\cite{mr93} could also wash out the difference between pair
winds and $e^-p$ disk winds. 
\subsection{Relativistic particles, cold and hot beams,
internal and external target photons}
By analogy with the local cosmic rays,
the {\bf relativistic particles} responsible for
the  $\gamma$-rays are expected to be electrons and protons.
In the tenuous plasma of the jets,
electrons (positrons)  produce $\gamma$-rays mainly by inverse-Compton
scattering or
synchrotron emission. For protons these processes
are generally unimportant owing to their larger mass.
The pion production on matter, which is important as a $\gamma$-ray
production process in the Galaxy, can be neglected due to the low
plasma density in radio jets.
However,  the photo-production of secondary pions and pairs in collisions
with ambient low-energy photons can be important.  
In spite of the hadronic cross-section $\sigma_{\rm p\gamma}=5\, 10^{-28}$~
cm$^2$ being much smaller than the Thomson cross-section $\sigma_{\rm T}=
6.7\, 10^{-25}$~cm$^2$, the proton cooling time scale can be as
short as the electron Thomson cooling time scale owing to 
the much larger inelasticity of  hadronic processes.
The proton and electron energy losses ($i=e,p$) are given by
\begin{equation}
-{dE_{\rm i}\over dt}={4\over 3}\sigma_{\rm T} \left(m_{\rm e}\over m_{\rm i}\right)^2
{B^2\over 8\pi}\gamma_{\rm i}^2f_{\rm i}c
\end{equation}
where the factors
$f_{\rm e}=1+{u_\gamma/ u_{\rm B}}$ and $f_{\rm p}=1+500{u_\gamma/
u_{\rm B}}$ take into account the relative contribution of 
synchrotron, inverse Compton, Bethe-Heitler pair production,
and pion production.
Balancing with energy gains due to Fermi acceleration 
\begin{equation}
{dE\over dt}=K e c^2 B
\end{equation} 
where $K<0.4$ depends on the diffusion coefficients
one obtains the maximum Lorentz factors
\begin{equation}
\gamma_{\rm p,max}=7\, 10^{10}K^{1\over 2}B^{-{1\over 2}}f_{\rm p}^{-{1\over 2}}
\end{equation}
for protons and
\begin{equation}
\gamma_{\rm e,max}=4\, 10^7 K^{1\over 2}B^{-{1\over 2}}f_{\rm p}^{-{1\over 2}}
\end{equation}
for electrons.  Adopting $K=0.1$, $B=1$~G, and $u_\gamma=u_{\rm B}$
this corresponds to maximum energies $E_{\rm p,max}=10^6$~TeV
and $E_{\rm e,max}=4$~TeV.  The assumption of $K=0.1$ is
certainly an overestimate for electrons which, owing to their
much smaller Larmor radius, are in resonance with much smaller scales of the 
expected inverse-power-law plasma turbulence spectrum than protons\cite{bs87}.  
The ratio of cooling time scales for
electrons and protons is given by
\begin{equation}
{t_{\rm e}\over t_{\rm p}}={1\over E_{\rm p}}{dE_{\rm p}\over dt}
\left({1\over E_{\rm e}}{dE_{\rm e}\over dt}\right)^{-1}=
\left(m_{\rm e}\over m_{\rm p}\right)^3{\gamma_{\rm p}\over \gamma_{\rm e}}
{f_{\rm p}\over f_{\rm e}} 
\end{equation}
from which one can readily
approximate the proton induced luminosity in terms of 
the electronic inverse-Compton and synchrotron luminosity
\begin{equation}
L_{\rm p}\approx
{u_{\rm p}\over u_{\rm e}}{t_{\rm e}\over t_{\rm p}}L_{\rm e}
={u_{\rm p}\over u_{\rm e}}\left(m_{\rm e}\over m_{\rm p}\right)^3
{\gamma_{\rm p}\over \gamma_{\rm e}}
{f_{\rm p}\over f_{\rm e}} L_{\rm e}
\end{equation}
where $u_{\rm p}$ and $u_{\rm e}$ denote the energy densities
in protons and electrons, respectively.
To estimate the ratio of Lorentz factors relevant for the
luminosities it is important to realize that
the synchrotron luminosity is mainly due to electrons with
$t_{\rm e}\le t_{\rm dyn}$ where $t_{\rm dyn}=r_{\rm sh}/c$ denotes the dynamical time scale
associated with the expansion of the shock front.
The  luminosity is not dominated by emission from the most
energetic electrons, but increases
only logarithmically above $t_{\rm e}= t_{\rm dyn}$
corresponding
to Lorentz factors 
$\gamma_{\rm e}\sim 10^2$ (infrared break). 
With $\gamma_{\rm p,max}\sim 10^9$  
one obtains  from Eq.(7) a maximum proton-induced luminosity
of the order of
\begin{equation}
L_{\rm p,max}\approx {u_{\rm p}\over u_{\rm e}} L_{\rm e}
\end{equation}
Photoproduction by ultra-relativistic protons in a
relativistic jet has been considered by a number of authors
assuming different locations of the $\gamma$-ray emitting
zone, either 
very close to the central black hole\cite{bl82,bk95},
at a distance of the order of the broad
emission line region radius\cite{p96}, or beyond the
broad line region where the jet is dominated 
by its own synchrotron radiation field\cite{mb92,m93}.
Interestingly, the proton maximum
energies and luminosities required to explain the blazar $\gamma$-rays
conspire to produce the observed flux of ultra-high energy cosmic
rays\cite{rb93}.  

The bulk flow of relativistic particles
can be distinguished as being either a {\bf cold beam} or a {\bf hot beam}.
In the former case, the particles have zero kinetic energies in the
frame co-moving with the bulk flow. The TeV $\gamma$-ray observations 
would then require cold beams with
particle energies of at least several TeV\cite{ckk92}.  
 In the latter case,
the bulk flow need only be moderately relativistic to accommodate for
the superluminal motion and the absence of pair attenuation, whereas
the co-moving frame
energies are ultra-relativistic.  Superluminal motion of radio knots
suggests bulk Lorentz factors  $\Gamma\sim10$ (at the $\sim 0.1$~pc
scale) and observations of the synchrotron turnovers indicate
co-moving frame Lorentz factors of up to several tens of GeV.
Cold beams could
be generated through ordered electric fields 
close to the central object\cite{bl82,bk95}  or electric drift
fields arising at standing shock waves\cite{bp96}.
Cold beams\cite{bl82,bk95,bkm96,dl97} are expected to slow down 
reaching bulk Lorentz factors of  $\Gamma\sim 10$ owing to
Compton drag\cite{p87,mk89}.
 Moreover, 
the same target radiation field which is assumed to be
up-scattered to $\gamma$-ray energies would
attenuate the $\gamma$-rays by pair
production on a comparable time scale above photon energies of $m_{\rm e}c^2$.
An interesting idea\cite{bp96} is that  
cold sub-beams embedded in a larger scale jet
could be produced  at
shocks in the winds of stars orbiting through the jet plasma.
To explain the variability in Mrk421, one would, however, need an unexpected
high density of mass-loosing stars in the central stellar cluster.
Since the Compton-drag on cold beams is difficult to avoid
and the bulk Lorentz factors inferred from radio observations
are much smaller than the Lorentz factors of the emitting particles,
one needs {\em in situ} particle
acceleration further downstream the jet as assumed
in the hot beam models.
Hot beam models with only one zone of emission in the jet
(a `blob') and {\rm external photons}
as the dominant target for inverse-Compton
scattering (external Compton models: EC)
have been published assuming that the blob is
close to the nucleus\cite{ds93} at a distance of
$(10^2-10^3)r_{\rm S}$ or at the distance of the broad emission
line region \cite{sbr94}
at $(10^3-10^4)r_{\rm S}$. In the
hadronic variant of this idea\cite{p96},
protons are assumed as the primaries initiating synchrotron
cascades by scattering external photons (E-PIC).
In  synchrotron-self-Compton models\cite{mgg92,gmd96,mt96}, the $\gamma$-rays
originate in the jet where it is dominated by the internal
synchrotron photons, i.e. presumably at distances of 
$>10^4r_{\rm S}$.  Again, there is a hadronic analogue,
the proton-initiated
synchrotron cascade model\cite{mb92,m93} 
(PIC, denoted as S-PIC in Fig.~4).
The closer
the $\gamma$-ray emitting zone to the center, the lower the $\gamma$-ray
energy which can be emitted without attenuation.
The inhomogeneous EC
models\cite{bl95,mhp95,rl97}
consider the emission of a Poynting flux/pair jet from close to the nucleus
to the edge of
the broad emission line region, the effect of this energy-dependent
$\gamma$-photosphere is that with increasing photon energy, the
$\gamma$-rays come from increasingly distant parts of
the jet.  
Recently, it has been
pointed out\cite{sm96} that the  
high density of  pairs required in these models at $r<10^2r_{\rm S}$
would lead
to a bump in X-rays from inverse-Compton scattering of disk
photons. Instead of a bump, blazars show a deep gap in X-rays.
As can be seen from Fig.~1, all possible
combinations of the basic assumptions have been investigated.
In Sect.~2,
the model predictions are confronted with observations.  
Section 3 concludes with a critical discussion
of the models for high-energy emission from blazars.
%\begin{figure}
%\centerline{\psfig{figure=picture.ps,width=9cm,height=9cm}}
%\caption[]{Schematic diagram of an AGN with the various assumed
% locations of the $\gamma$-ray emitting zones in the jet.  Beyond
%$\sim 10^5r_{\rm S}$ internal synchrotron photons
%are commonly believed to be dominant.}
%\end{figure}

\section{Key observational results and model predictions} 
 
\subsection{Fraction of $\gamma$-ray sources among blazars}

A fraction of $\sim 10$\% of all blazars have been detected\cite{mea97} in the 
EGRET pass-band 100 MeV -- 10 GeV.  
The {\rm blazar fraction} refers to those flat-spectrum radio sources 
which would have been above the EGRET flux sensitivity if the ratio 
between their $\gamma$-ray and 5 Ghz radio fluxes were 
at least as large as the lowest value for the EGRET detected 
blazars. 
An obvious explanation of the 10\% would be flux variability which is, 
however, inconsistent
with the $30$\% duty cycle of the strong EGRET detections. 
Intermittent behavior is suggested by 
interferometric radio observations which show 
months or years between the ejection of new radio knots, 
possibly associated with periods of $\gamma$-ray flaring 
and quiescence.    Nevertheless, some quasi-stationary
emission models make 
definite predictions about the $\gamma$-ray blazars representing a true
subset of all blazars.  EC models predict 
a stronger Doppler boosting 
of the $\gamma$-ray flux $\propto \delta^{4+2\alpha}$ compared with 
the radio-to-optical synchrotron flux which is  
boosted only by the factor $\delta^{3+\alpha}$.    With
$\alpha=1$, the prominence 
of the $\gamma$-ray component increases  $\propto\delta^2$ 
implying that the most pronounced $\gamma$-ray blazars must be the ones 
with the largest Doppler factors (smallest viewing angles or  
largest bulk Lorentz factors).  This prediction can be tested  
experimentally with interferometric radio observations. 
There are claims that, indeed, the strong EGRET sources have 
atypically high values of the Doppler factors\cite{amea96},  
whereas there the Doppler factors of the EGRET sources among the 
S5 radio sources are normal for blazars\cite{kea96}. 
In the PIC model, the ratio between $\gamma$-ray and radio-to-optical 
luminosity depends on the proton-to-electron ratio and on the 
ratio of cooling time scales, see Eq. (5).    There is no reason to  
believe that either one should be constant, it is rather plausible 
that the numbers vary with 
shock obliquity, Lorentz factor, and jet magnetization.   
Proton
acceleration is likely to be
limited due to the particle gyroradius  $r_{\rm L}$ 
exceeding the  shock radius of curvature $r_{\rm sh}$
rather than due to energy losses.
If this is indeed the case, one obtains
$r_{\rm L}\propto r_{\rm sh}\propto r_{\rm j}$  at the maximum energy 
with $r_{\rm j}$ denoting the jet radius.  From eq.(7),  
the ratio of proton induced $\gamma$-ray  
and electronic synchro/Compton luminosity is then given by 
$L_{\rm p}/L_{\rm e}\propto 
\gamma_{\rm p,max}\propto r_{\rm j}B$.  Jet formation models 
give the scaling $r_{\rm j}\propto M_{\rm bh}$ and $B\propto l_{\rm edd}^{1/ 2} 
M_{\rm bh}^{-{1/2}}$ 
(from $u_{\rm B}=u_{\rm edd}$). The parameter $l_{\rm edd}=L/L_{\rm edd}$ 
denotes the luminosity in units of
the Eddington luminosity.   Hence one 
obtains $L_{\rm p}/L_{\rm e}\propto M_{\rm bh}^{1/ 2}l_{\rm edd}^{ 
1/ 2}$.  Thus, in the PIC model quasars, which are intrinsically 
more luminous than BL Lacs and therefore require larger $M_{\rm bh}$'s or 
$l_{\rm edd}$'s, are expected to be 
relatively more luminous in $\gamma$-rays 
than BL Lacs which is consistent with observations.
The non-detected blazars would correspond to jets
associated with  sub-Eddington  accretion flows.

\subsection{Amplitude of $\gamma$-ray/radio luminosity ratio}
 
The ratios between the observed $\gamma$-ray and radio luminosities
range roughly between $10^2$ and $10^4$ (Fig.~2). If $\gamma$-ray
blazars have a true distribution of   
$\gamma$/radio amplitudes 
larger than 100, so that only the 10\% with the largest amplitudes
show up above
the EGRET sensitivity threshold, the observed blazar fraction 
could easily be explained.
In the EC model, the amplitude 
distribution is triggered by the jet Doppler 
factor and with $\delta=1- 10$  
the model predicts amplitudes between $\delta^2=1-100$. 
Maximum Lorentz factors somewhat larger than 10 would therefore be 
sufficient to explain the width of the amplitude distribution. 
As shown above, 
the PIC model requires that the product of black hole mass 
and Eddington ratio spans a range of more than $10^4$ which is  
also necessary to explain the range of blazar bolometric 
luminosities.  Whereas the EC and PIC models entail 
satisfactory explanations of the blazar fraction and the amplitude 
distribution, the SSC models are inconsistent with the observations 
of a few high-luminosity quasars such as 3C279 and PKS 0528-134 
in which the $\gamma$-ray luminosity exceeds the radio-to-optical 
synchrotron luminosity by factors $10-100$.  This follows from 
re-writing the ratio of the singly scattered Compton luminosity 
to the synchrotron luminosity in terms of the Thomson optical 
depth\cite{s96} 
which yields $L_{\rm SSC}/L_{\rm syn}\approx 
\tau_{\rm T}\ln[\nu_{\rm c}/\nu_{\rm b}]\approx 5\tau_{\rm T}$ for the 
typical synchrotron 
spectra with spectral index $\alpha=1$ between  
frequencies $\nu_{\rm b}$ in the infrared and $\nu_{\rm c}$ in the 
ultraviolet.  This does not say that the SSC mechanism is 
generally unimportant, 
it could still be the dominant process in the BL Lac objects where 
$L_\gamma \approx L_{\rm syn}$.  

\begin{figure}
\vskip-0.5cm
\centerline{\psfig{figure=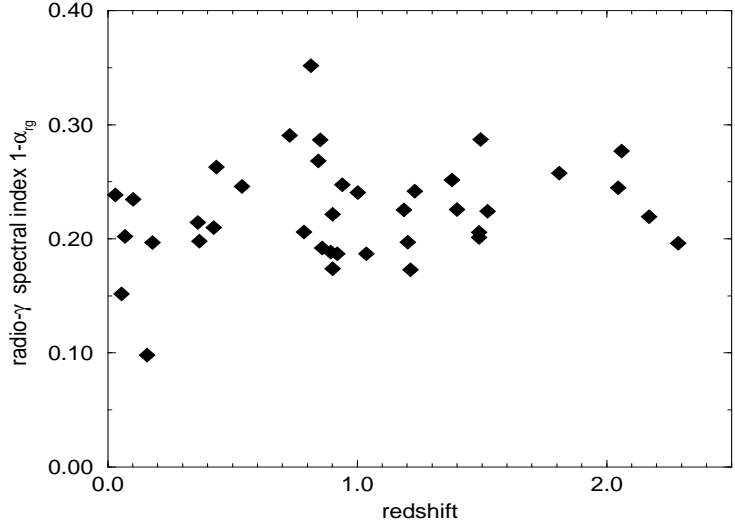,width=11cm,height=8cm}}
\caption[]{Redshift distribution of the 5~GHz to 100~MeV
spectral index $\alpha_{\rm rg}$ for the strong EGRET blazars
\cite{mea97}.
The $\gamma$/radio luminosity ratio is given by
$L_\gamma/L_{5}=(5\, 10^{12})^{1-\alpha_{\rm rg}}$.}
\end{figure}

\subsection{Spectrum}

All models do
rather well in fitting spectral data  (for an example, see Fig.~3).
However, the sharp MeV break seen in some of the
 spectra  
is reproduced by EC and SSC models only by the {\em ad hoc} assumption of
a break in the electron spectrum and it is difficult
to find a physical process which could lead to this particular
break.  Coulomb losses must be negligible, otherwise the 
density of electrons is too large to comply with the weakness
of the X-ray emission (absence of Sikora-bump\cite{s96}).
Inhomogeneous EC models\cite{bl95,mhp95} can
easily reproduce MeV breaks
assuming steep gradients along the jet. In the PIC model, 
the high-energy emission is due
to {\em unsaturated} synchrotron cascades
which, contrary  to {\em saturated} cascades,
can produce very hard X-ray spectra depending on the slope
of the target photon spectrum.
However, $\gamma$-ray attenuation is required to
keep the  $\gamma$-ray flux below  Whipple limits.
A natural source of the attenuating target photons is a heated dust torus.
The PIC model predicts an average spectrum $\propto E^{-2}$
from MeV to TeV, and roughly $\propto E^{-3}$ above\cite{mea96}.  
 The upper break is less certain, it
depends on the range of distances in which
proton cooling is important.
In the PIC model, $\gamma$-ray emission is considered only from the
part of the jet where the infrared target photons become optically
thin. This is expected to be the most $\gamma$-ray luminous part of the jet.
%\vskip-4cm
\begin{figure}
\vskip-1.5cm
\centerline{
\psfig{figure=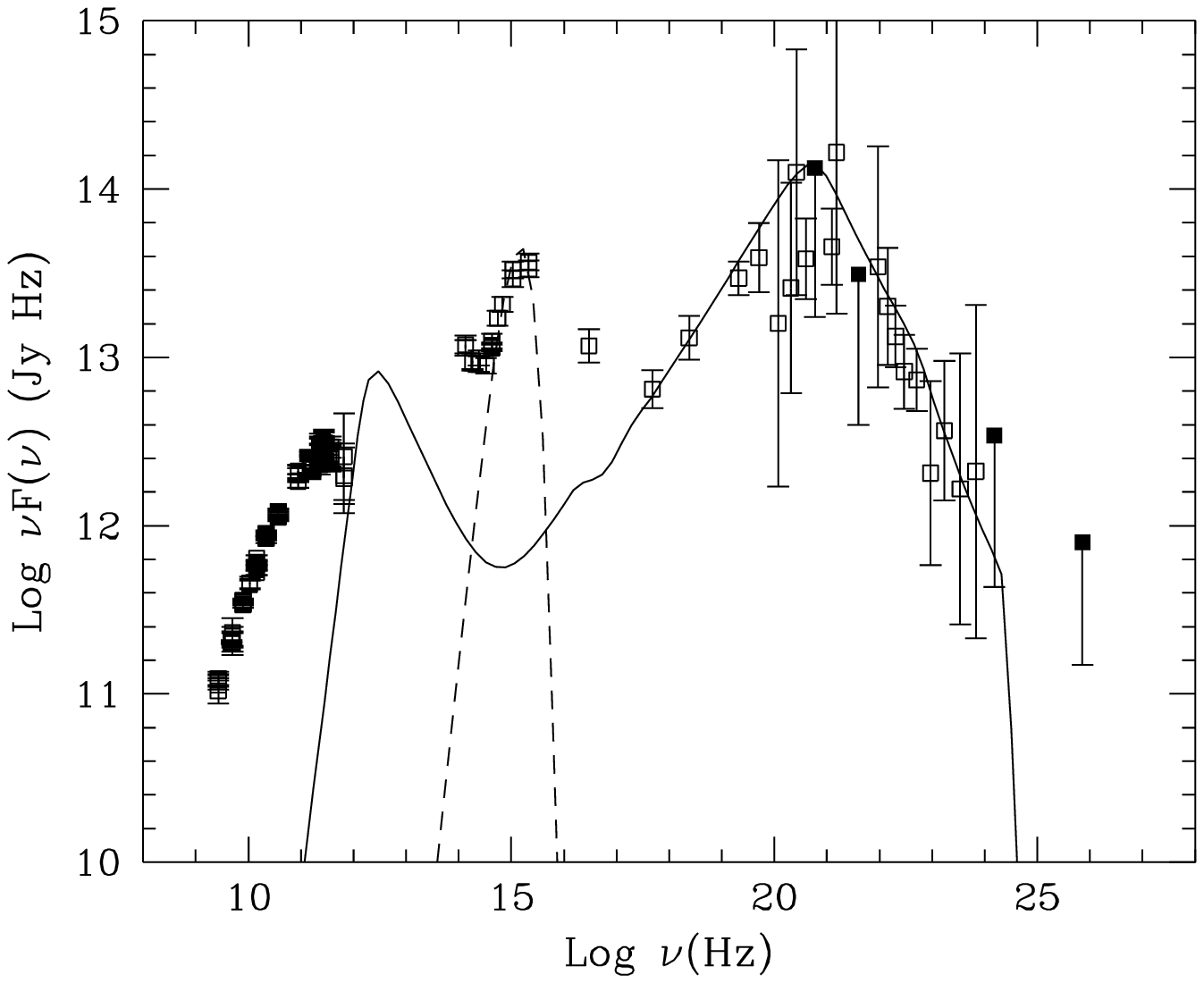,width=8cm,height=10cm}\hskip-2cm
\psfig{figure=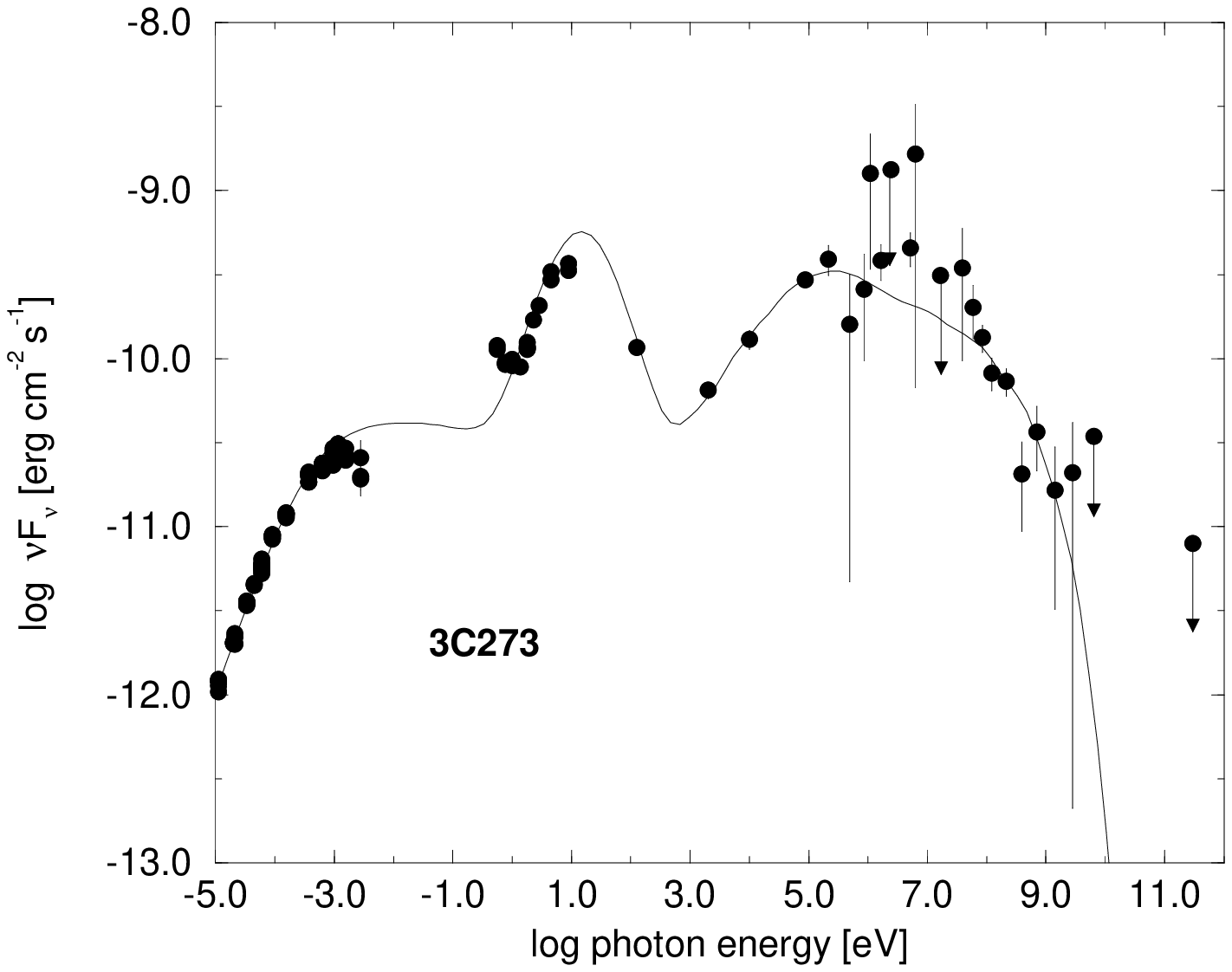,width=8cm,height=10cm}
}
\vskip-5cm
\centerline{\psfig{figure=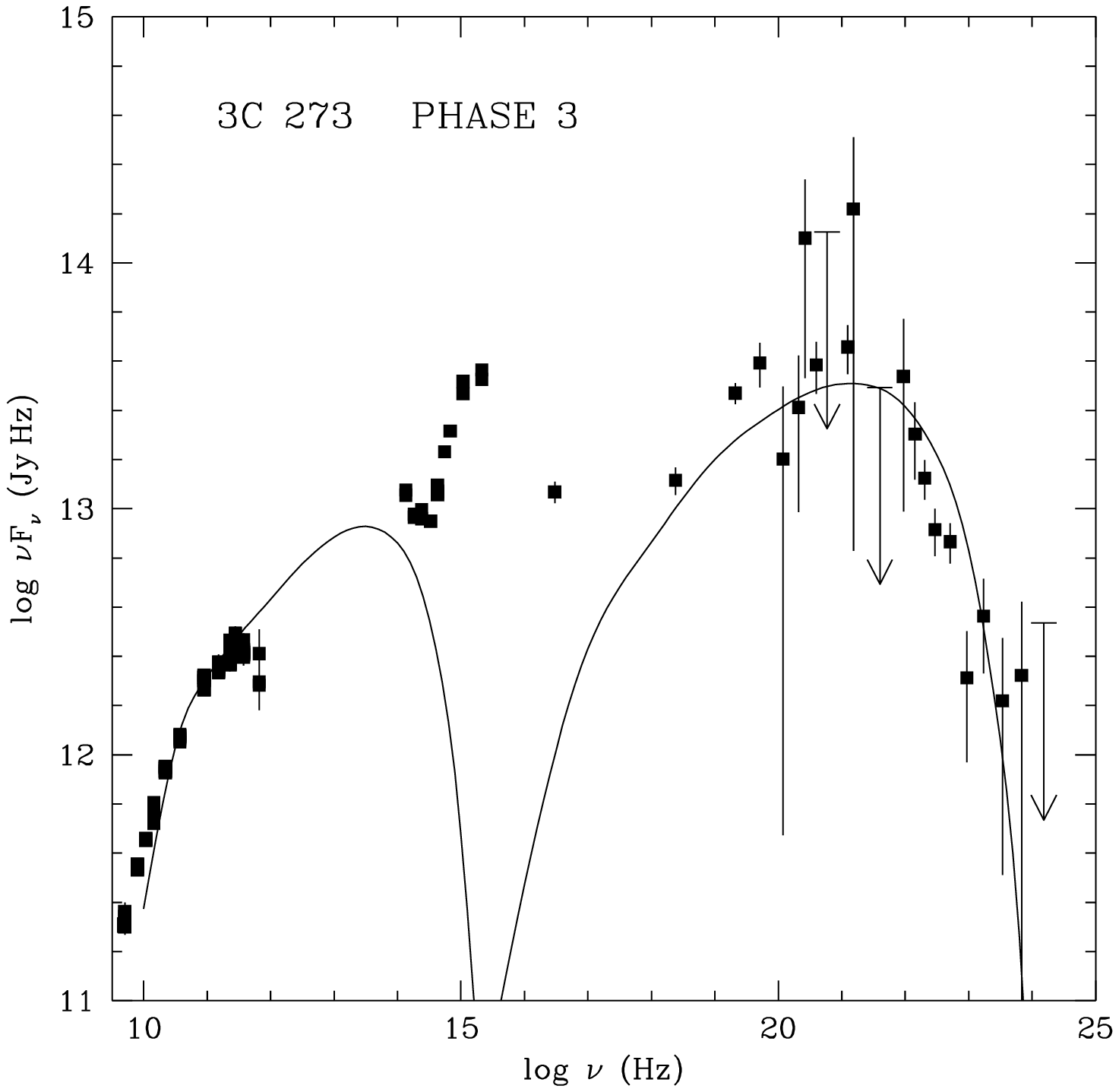,width=7cm,height=8cm}}
\vskip-3cm
\caption[]{EC, PIC, and SSC model fits to 
3C273 quasi-simultaneous multi-frequency
spectrum\cite{vmea97}}
\end{figure}

\subsection{Cut-off}

Since the $\gamma$-ray spectra are broad 
featureless continua, the cut-off energy is of great
diagnostic value.  No cut-off is generally seen in the $\gamma$-ray
spectra of blazars up to $\sim 10$~GeV above which the EGRET sensitivity
drops rapidly.  Generally, the flux level of the EGRET spectra
extrapolated to air-Cerenkov energies exceeds the sensitivity of
existing telescopes.  Nevertheless, only 2(3) of the nearest blazars
have been detected at TeV energies\cite{w97} which is commonly ascribed to the attenuation of $\gamma$-rays
due to pair production in the extragalactic infrared background.  
The large zenith angle
observations\cite{kr97} of Mrk421  indicate
that the spectrum extends at least to 5-8~TeV with no evidence
for attenuation.   
A possible excess of $\gamma$-ray showers at still higher energies
from the stacked source positions of the nearest blazars 
(including Mrk421) with
redshifts $z\le 0.07$ has been claimed\cite{mw96}
on the basis of an advanced analysis of HEGRA array data.
The significance of this excess 
has not yet been established unequivocally \cite{ah96}. 
Eqs.(4) and (5) show a quite general result regarding the maximum
possible energy in leptonic and hadronic emission models.
If the observed 5-8~TeV emission from Mrk421 is to be explained
by any model in which the primaries are accelerated electrons,
the electron Lorentz factors must reach at least $10^7$. 
The same electrons are assumed to produce the radio-to-soft-X-ray
spectrum by synchrotron emission.
With an observed turnover energy of $\epsilon\sim 300$~eV and 
the formula for the characteristic synchrotron
energy $\epsilon\sim
10^{-8} B\gamma^2$~eV one obtains $B\sim 3\, 10^{-4}$~G.
Eq.(5)  implies that the magnetic field in the acceleration
region must be $< 1$~G
adopting (optimistically) $K=0.1$ and $f_{\rm e}=2$
which is consistent with the above estimate.
However, fields with a strength of $B\sim 10^{-4}$~G are
found observationally in the hot spots of 
jets at the {\em kiloparsec} scale\cite{hcp94}.  
This is at least $\sim 10^6$ times larger
than the size of the $\gamma$-ray zone inferred from variability.
Adiabatic flux compression to the scale of the $\gamma$-ray
emitting zone yields
$B$-fields somewhere in the range $0.1-100$~G.
Taking into account a Doppler factor $\delta\sim 10$ does not
change these conclusions significantly.
Therefore, the magnetic field values implied by SSC models fitted
to $\sim 10$~TeV emission
seem unrealistically low.  An estimate similar to
Eq.(5) yielding $\sim 10$~TeV
as the maximum possible energy for leptonic models
has been calculated\cite{bkm96}  assuming ordered
electric fields. 
The leptonic limit is in marked
contrast to the hadronic models which naturally produce
$>10$~TeV energies.
 
\subsection{Variability}

Doubling time scales of days to months seem to be
common among  strong EGRET sources\cite{vmea95}.  Considering
the long
integration times required to obtain 5-$\sigma$ detections
the minimum variability time scales could also be shorter.
Much shorter time scale variability has been discovered in
Mrk421 at TeV energies taking advantage of the superior
aperture of the air Cerenkov method\cite{bea96}. 
The shortest time scales expected for emission from  
blazar jets are of the order of
$\Delta t\sim {r_{\rm j}/( c\delta)}$ where the
minimum jet radius can be estimated from
asymptotic solutions of the equations
for the radial structure of self-collimating 
hydromagnetic jets\cite{ac93}.
This yields $r_{\rm j}=10\Gamma r_{\rm S}=3\, 10^{15}\Gamma_{10}m_8$ cm 
where $\Gamma_{\rm 10}=\Gamma/10$ and $m_8=M/10^8M_\odot$ 
denote bulk Lorentz factor and
the mass of the black hole, respectively, so that one obtains
\begin{equation}
\Delta t_{\rm min}\sim {10 \Gamma  r_{\rm S}\over c\delta}\sim 5\, 10^3 m_8~{\rm s}
\end{equation}
assuming that blazars have favorably large Doppler factors $\delta\sim\Gamma$.
In principle, one could further reduce this minimum time scale by
another $\delta^{-1}$ factor if the jet opening angle exceeds $\sim 1/\Gamma$
or if the emission region is very thin and running through inhomogeneities.
In spite of the short variability time scales, the pair creation
optical depth of the emission region can be very small.
The maximum optically thin energy for
a spherical blob is given by
$E\sim 3 \delta_{10}^6L_{44}^{-1}[\epsilon/\delta^2]\Delta t_4$~TeV
where $L_{44}$ denotes the apparent luminosity in  units of
$10^{44}$~erg/s and $\Delta t_4$ the variability time scale in units
of $10^4$~s.  The low compactness is an
essential requirement for the PIC model which would otherwise overproduce
X-rays.   In fact, the PIC model 
has predicted emission above TeV from the
presence of X-ray gaps in the blazar spectra\cite{m93}.
 
\subsection{Correlations} The claimed $\gamma$-ray/optical
correlation\cite{w96} is consistent with EC models, if the variations are due to a varying
bulk Lorentz factor.  In this case, one would expect the amplitude of the
$\gamma$-ray variations to be larger than the synchrotron variations, as seems
to be the case in 3C279\cite{mea94}.  SSC models predict
proportional variations in this case and quadratic ones, if the variations are
due to a variable comoving frame maximum Lorentz factor.  The latter is
suggested by UV/X-ray observations of blazars.  The PIC
model predicts basically the same variability pattern as the SSC model (protons
instead of electrons scattering off synchrotron photons).  The only difference
is that variations of the electron maximum energy do not {\em necessarily} imply
the same variations of the proton maximum energies, if the maximum energies are
determined by a subtle balance between acceleration, energy losses, and
expansion.  Thus, the PIC model is less predictive in this respect.
The inhomogeneous EC models, in which the time scales for
$\gamma$-ray variations increase with energy and in which the TeV and optical
emission sites are not in the same part of the jet, are clearly inconsistent
with the claimed correlations.

\subsection{Warm dust $\gamma$-ray attenuation}

As outlined in Sect.~1, the $\gamma$-ray emission models 
assuming external photons as the dominant target rely on
a parametrization of the inner parsec of AGN as given
by unified schemes.  An element of the unified schemes 
which has more recently been recognized to be
important\cite{pb96} is the 
dust torus heated by nuclear radiation feeding the accretion
flow onto the black hole.
The sublimation temperature of dust  
is $\sim 1700$~K and determines the inner edge of
the torus to be at
\begin{equation}
r_{\rm d}\sim 2\, 10^4 \left(l_{\rm edd}\over m_8\right)^{1\over 2}
r_{\rm S} 
\end{equation}
assuming a covering fraction of 0.5.
This is roughly the same as the radius of the broad line region.
The thermal NIR photons emitted from the warm dust torus
absorb practically any $\gamma$-ray with energy above $\sim 300$~GeV
emitted from $r<r_{\rm d}$.  On the basis of a straw person's
unified scheme, 
EC models  are therefore ruled out
as an explanation of TeV sources. 
 
\begin{figure}
\vskip-2cm
\centerline{\psfig{figure=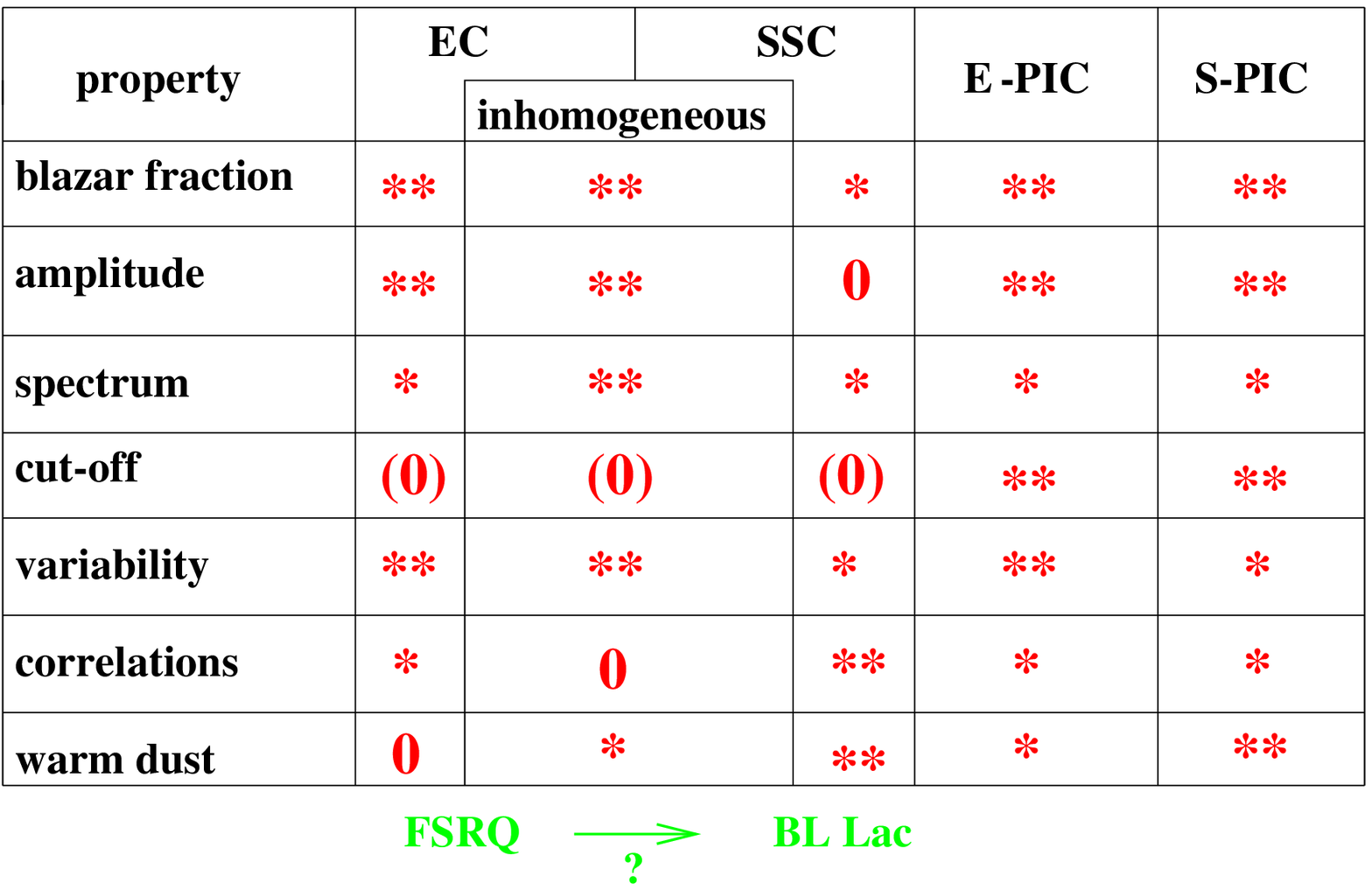,width=10cm,height=10cm}}
\caption[]{Tentative score chart (model explains observations
naturally [**], with ad hoc assumptions [*], not at all [0]).
Blazar emission above $\sim 10$~TeV would practically rule
out the leptonic models, as indicated by [(0)]
}
\end{figure}

\section{Discussion and conclusions}
The results of Sect.~2 are summarized in Fig.~4.
More refined models with more parameters could certainly
improve some of the low scores, but some fundamental problems
remain.  The EC models have difficulties explaining the
TeV emission due to the attenuating warm dust radiation field
and the SSC models have difficulties explaining the enormous
$\gamma$-ray luminosities of some quasars.  The very attractive
inhomogeneous EC models fail to explain $\gamma$-ray/optical
correlations.
The problems of the leptonic models could be ameliorated by
assuming a transition from SSC to EC behavior when going
from BL Lacertae objects to flat-spectrum radio quasars. 
Even if the emission properties can be explained by
electron acceleration only, the absence of the Sikora X-ray bump
is more in line with the energy transport
in jets being dominated by protons rather than pairs and Poynting flux.

In any model, the observed short variability time scales
observed in Mrk421 push
the required acceleration times toward their limiting values allowed
by diffusive shock acceleration 
and indicate that the size of the $\gamma$-ray
emitting region is much smaller than its distance to the
active nucleus. This is
expected if the emission originates in the radiative downstream regions of  
shocks  with
a thickness of the order of the optical synchrotron cooling
length. 

Observations above 10~TeV are crucial for
the discrimination between leptonic and hadronic
models, but they are unfortunately hampered by the 
expected cosmic attenuation of $\gamma$-rays at these
energies even for nearby blazars.  The detection 
of Mrk421 at 5~TeV and the possible cumulative excess
of HEGRA events from the nearest blazars  
are  encouraging hints that the cosmic transparency  
is not exceedingly strong.  
Ultimately,  observations
with high-energy neutrino telescopes\cite{z97} could 
bring us closer to 
an understanding of the puzzling nature of blazars and 
their relation to ultrahigh-energy cosmic rays.

\section*{Acknowledgments}
Support by a European Training and Mobility of
Researchers grant is greatly acknowledged.


\section*{References}
\begin{thebibliography}{99}
\bibitem{ah96}
F.A. Aharonian,  G. Heinzelmann (HEGRA collaboration),
Summary of the HEGRA reports given at the  
{\em $15^{\rm th}$ European Cosmic Ray Symposium
1996, Perpignan, France, 26-30 August 1996}, astro-ph/9702059
%\bibitem{}
%R. Antonucci, {\em Ann. Rev. A.\&A.} {\bf 31}, 473 (1993)
\bibitem{ac93}
S. Appl, M. Camenzind, \Journal{\AA}{274}{699}{1993}
\bibitem{bp96}
W. Bednarek, R.J. Protheroe, \Journal{\MNRAS}{}{submitted}{1996}
\bibitem{bkm96}
W. Bednarek, J.G. Kirk, A. Mastichiadis, \Journal{\AA}{307}{L71}{1996}
\bibitem{bk95}
W. Bednarek, J.G. Kirk, \Journal{\AA}{294}{366}{1995}
\bibitem{bs87}
P.L. Biermann, P.A. Strittmatter, \Journal{\APJ}{322}{643}{1987}
\bibitem{bl95}
R.D. Blandford, A. Levinson, \Journal{\APJ}{441}{79}{1995}
\bibitem{bz77}
R.D. Blandford, R.L. Znajek, \Journal{\MNRAS}{179}{433}{1977}
\bibitem{bea96}
H.H. Buckley, et al., \Journal{\APJ}{472}{L9}{1996}
\bibitem{bl82}
M.L. Burns, R.V.E. Lovelace, \Journal{\APJ}{262}{87}{1982}
\bibitem{c86}
M. Camenzind, \Journal{\AA}{162}{32}{1986}
\bibitem{ckk92}
P.S. Coppi, J.F. Kartje, A. K\"onigl, {\rm BAAS} {\bf 24}, 732 (1992)
\bibitem{dl97}
A. Dar, A. Laor, \Journal{\APJ}{478}{L5}{1997}
\bibitem{ds93}
C.D. Dermer, R. Schlickeiser, \Journal{\APJ}{416}{458}{1993}
\bibitem{gmd96}
G. Ghisellini, L. Maraschi, L. Dondi, \Journal{\APJS}{120}{503}{1996}
\bibitem{hcp94}
D.E. Harris, C.L. Carilli, R.A. Perley, {\em Nature} {\bf 367}, 713 (1994)
\bibitem{k97}
G. Kanbach, this volume (1997)
\bibitem{kr97}
F. Krennrich, this volume (1997)
\bibitem{kea96}
T. Krichbaum, et al., in: {\em Proc. of the Heidelberg
Workshop on $\gamma$-ray Emitting AGN},
MPI Preprint, MPI H-V37-1996, p.~97 (1996)
\bibitem{mea96} 
K. Mannheim, S. Westerhoff, H. Meyer, H.-H. Fink, \Journal{\AA}{315}{77}{1996}
\bibitem{m93}
K. Mannheim, \Journal{\AA}{269}{67}{1993}
\bibitem{mb92}
K. Mannheim, P.L. Biermann, \Journal{\AA}{253}{L21}{1992}
\bibitem{mea94}
L. Maraschi, et al., \Journal{\APJ}{435}{L91}{1994}
\bibitem{mgg92} 
L. Maraschi, G. Ghisellini, A. Celotti, \Journal{\APJ}{397}{L5}{1992}
\bibitem{mhp95}
A. Marcowith,  G. Henri, G. Pelletier, \Journal{\MNRAS}{277}{681}{1995}
\bibitem{mt96}
A.P. Marscher, J.P. Travis, \Journal{\APJS}{120}{537}{1996}
\bibitem{amea96}
A.P. Marscher, et al., in: {\em Proc. of the Heidelberg
Workshop on $\gamma$-ray Emitting AGN},
MPI Preprint, MPI H-V37-1996, p.~103 (1996)
\bibitem{mea97}
J.R. Mattox, J. Schachter, L. Molnar, R.C. Hartman,
A.R. Patnaik, \Journal{\APJ}{481}{May 20 issue}{1997}
\bibitem{mk89}
F. Melia, A. K\"onigl, \Journal{\APJ}{340}{162}{1989}
\bibitem{mr93}
P. M\'esz\'aros, Rees M.J., \Journal{\APJ}{405}{278}{1993}
\bibitem{mw96} 
H. Meyer, S. Westerhoff, in: {\em Proc. of the Heidelberg
Workshop on $\gamma$-ray Emitting AGN},
MPI Preprint, MPI H-V37-1996, p.~39 (1996)
\bibitem{vmea97}
C. von Montigny, et al., \Journal{\APJ}{}{in press}{1997}
\bibitem{vmea95}
C. von Montigny, et al., \Journal{\APJ}{440}{525}{1995}
\bibitem{p97}
P. Padovani, this volume (1997)
\bibitem{p87}
E.S. Phinney, in: {\em Superluminal radio sources, Proceedings
of the Workshop, Pasadena, CA, Oct 28-30, 1986}, p. 301 (Cambridge University
Press, 1987)
\bibitem{p96}
R. Protheroe, to appear in: {\em Accretion Phenomena and Related Outflows},
IAU Colloq. 163, ed. D. Wickramashinge et al. (1996)
\bibitem{pb96}
R. Protheroe, P.L. Biermann, {\em Astropart. Phys.} {\bf 6}, 45 (1996)
\bibitem{rb93}
J.P. Rachen, P.L. Biermann, \Journal{\AA}{273}{377}{1993}
\bibitem{rl97}
M.M. Romanova, R.V.E. Lovelace, \Journal{\APJ}{475}{97}{1997}
\bibitem{s96}
R. Schlickeiser, in: {\em Proc. of the Heidelberg
Workshop on $\gamma$-ray Emitting AGN},
MPI Preprint, MPI H-V37-1996, p.~147 (1996)
\bibitem{sm96}
M. Sikora,  G. Madejski,  in: {\em Proc. of the Heidelberg
Workshop on $\gamma$-ray Emitting AGN},
MPI Preprint, MPI H-V37-1996, p.~153 (1996)
\bibitem{sbr94}
M. Sikora, M.C. Begelman, M.J. Rees, \Journal{APJ}{421}{153}{1994}
\bibitem{w96}
S.J. Wagner, \Journal{\APJS}{120}{495}{1996}
\bibitem{w97}
T. Weekes, this volume (1997)
\bibitem{wc95}
A.S. Wilson, E.J.M. Colbert, \Journal{\APJ}{438}{62}{1995}
\bibitem{z97}
E. Zas, this volume (1997)



\end{thebibliography}
\end{document}